\documentclass[aps,prx,twocolumn]{revtex4-2}

\usepackage{graphicx}
\usepackage{xcolor}
\usepackage{amsmath}
\usepackage{amssymb}
\usepackage{siunitx}
\usepackage[english]{babel}

\usepackage[colorlinks=true, pdfstartview=FitV, linkcolor=blue,
citecolor=blue, urlcolor=blue]{hyperref}
\usepackage[capitalise]{cleveref}

\begin{document}
\title{Electron-beam lithography of nanostructures at the tips of scanning probe cantilevers}
\author{L. Forrer$^{1,2}$, A. Kamber$^{1}$, A. Knoll$^{3}$, M. Poggio$^{1,2}$, and F. R. Braakman$^{1,2,\ast}$}

\affiliation{\vspace{1em} 1: Department of Physics, University of Basel, 4056 Basel, Switzerland,}
\affiliation{ 2: Swiss Nanoscience Institute, University of Basel, 4056 Basel, Switzerland}
\affiliation{3: IBM Research Europe - Z\"{u}rich, 8803 Rueschlikon, Switzerland}
\date{\today}

\begin{abstract}\noindent We developed a process to fabricate nanoscale metallic gate electrodes on scanning probe cantilevers, including on the irregular surface of protruding cantilever tips. The process includes a floating-layer technique to coat the cantilevers in electron-beam resist. We demonstrate gate definition through a lift-off process, as well as through an etching process. The cantilevers maintain a high force sensitivity after undergoing the patterning process. Our method allows the patterning of nanoscale devices on fragile scanning probes, extending their functionality as sensors.
\end{abstract}
\maketitle
\section*{Introduction}
\noindent Cantilevers are widely used in scanning probe microscopy, where they find application as force transducers in e.g. atomic force microscopy, magnetic force microscopy, and kelvin probe microscopy~\cite{Meyer2021}. These cantilevers are usually optimized for mechanical properties~\cite{Giessibl2003}, and include only simple tip design features. Integration of more advanced sensors on cantilever tips would allow the combination of extremely sensitive detectors such as superconducting quantum interference devices~\cite{Wyss2022}, single-electron transistors and quantum dot devices~\cite{Yoo1997, Khivrich2020}, and other quantum sensors with the navigational toolbox that scanning probe microscopy cantilevers provide. Such advanced devices typically require patterning of multiple metallic or superconducting gate and contact electrodes of nanoscale dimensions. 

\noindent Electron beam lithography (EBL) is a widespread technique to pattern nanoscale structures, such as gate electrodes. EBL can achieve high patterning resolution, in practice resulting in smallest feature sizes of $\sim$\SI{10}{\nano\meter}~\cite{Vieu2000}. Moreover, EBL enables batch production and many standard recipes have been previously developed, facilitating a fast fabrication development process and high device yield. Typically, in EBL, a planar substrate is first spin-coated with a resist, which is subsequently irradiated with electrons, developed, and processed further. However, spin-coating of small samples and substrates with non-flat surfaces does not provide a resist layer of acceptable uniformity, due to effects such as edge-beading, lift-off problems, and imperfect pattern transfer. Nonetheless, there are many applications where EBL on such irregular substrates would be of interest. 

\noindent Various other techniques of coating such substrates with a resist layer exist, including spray coating~\cite{Linden2011}, the use of ice resists~\cite{Han2012}, evaporative methods~\cite{Zhang2014}, and resist transfer techniques~\cite{Zhou2000, Chang2014, Nilsen2019}. Furthermore, focused-ion or electron-beam deposition~\cite{Jaafar2020} and milling methods can be used to pattern irregular substrates. 
However, these techniques are either slow, expensive, only demonstrated for photoresist, or are not tailored towards patterning on fragile substrates, such as scanning probe microscopy cantilevers, which are vulnerable suspended structures with a length typically of the order of \SI{100}{\micro\meter} and width and thickness of only a few \SI{}{\micro\meter}.

\noindent Based on an approach used for patterning on flakes of 2D materials~\cite{Leis2021}, we developed a fabrication process to pattern nanoscale structures on prefabricated cantilevers through EBL. The process involves coating the cantilever with an electron-beam resist using a floating-layer method. We fabricate metallic gates on various cantilever geometries, including on sloped and tipped parts of the cantilevers. We demonstrate two methods of gate definition: a lift-off process and an etching process. By employing either a standard electron-beam irradiation dose or a much higher dose, we use a poly(methyl methacrylate) (PMMA) layer either as a (standard) positive resist for the lift-off process, or as a negative resist for the etching process. The process allows us to produce gate arrays with gate width down to \SI{70}{\nano\meter} and pitch down to \SI{90}{\nano\meter}. We have successfully applied the method to pattern different types of cantilevers with good reproducibility.

\section*{Floating PMMA method}
\begin{figure*}[t]
\includegraphics[width=0.96\textwidth]{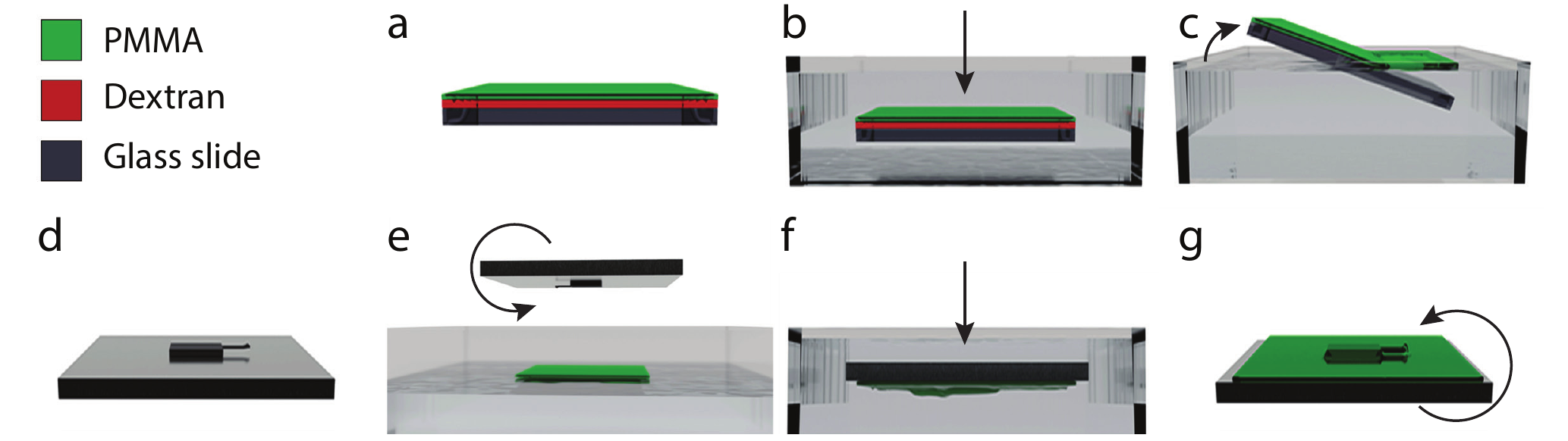}
\caption{Overview of resist-coating process. See main text for a step-by-step description of the process.
}
\label{fig:Figure1}
\end{figure*}

\noindent \cref{fig:Figure1} schematically shows the key steps of our floating-layer resist coating method. First, a \SI{15}{\percent} solution of dextran in deionized (DI) water is deposited onto a clean glass slide and spun at 2000\,rpm for \SI{40}{\second}. Afterwards, the slide is baked at \SI{150}{\degreeCelsius} for \SI{3}{\minute}. The desired resist layer is then spin-coated on top of the dextran film. For the work shown here, we employ as resist \SI{4.5}{\percent} PMMA with 950 K molecular weight dissolved in anisole (Allresist AR-P 672.045). The PMMA is spun at 4000\,rpm for \SI{40}{\second} and baked at \SI{150}{\degreeCelsius} for \SI{3}{\minute}, resulting in a resist layer thickness of approximately \SI{230}{\nano\meter} (\cref{fig:Figure1}a). Submerging the prepared slide into DI water for about \SI{15}{\minute} causes the dextran film to dissolve (\cref{fig:Figure1}b). After this, the slide is removed from the water, tilted by a small angle, and slowly inserted back into the water. This causes the PMMA to peel off the glass slide and float on the water surface, as illustrated in \cref{fig:Figure1}c.

\noindent Next, the desired sample is fixed onto a metal holder. In our case, we use a non-contact atomic force microscopy cantilever (NanoAndMore ATEC-NC) and fix it to an aluminum square slide with Kapton tape (\cref{fig:Figure1}d). We find that it is beneficial for the adhesion of the resist to the sample to perform UV-ozone plasma treatment for \SI{5}{\minute}. To coat the sample with the floating PMMA layer, it is turned upside down (\cref{fig:Figure1}e) and slowly lowered into the floating PMMA until the sample is fully submerged, as shown in \cref{fig:Figure1}f. The sample is then removed from the water at a small angle. To remove residual water, the sample is first dried with tissues, followed by a baking step at \SI{150}{\degreeCelsius} for \SI{3}{\minute}. This results in a sample which is coated by a uniform layer of PMMA (\cref{fig:Figure1}g), with minimal wrinkles or bubbles.

\noindent In the next sections, we use this method of resist transfer to demonstrate EBL patterning of gate electrodes using a lift-off process, as well as using an etching process. The lift-off process is an additive technique, where a thin metallic layer is deposited on top of the patterned and developed positive resist. In a next step, the resist is washed away in a solvent, leaving deposited material only in the areas where the resist was exposed. This technique is useful when etching or metallization can damage parts of the substrate material, for instance when producing electrical contacts to semiconductor nanostructures or fragile 2D materials. 
The etching process is a subtractive technique, where a thin metallic layer is deposited onto the sample prior to deposition of the resist. The negative resist is then patterned and developed, after which the deposited layer is etched away only where the resist is not exposed. Such an etching process avoids problems associated with lift-off, such as retention of residual resist and deposited material in unwanted areas of the substrate, and can produce finer structures than possible through lift-off.
\section*{Gates on cantilevers - lift-off process}
\begin{figure}[b]
	\includegraphics[width=0.48\textwidth]{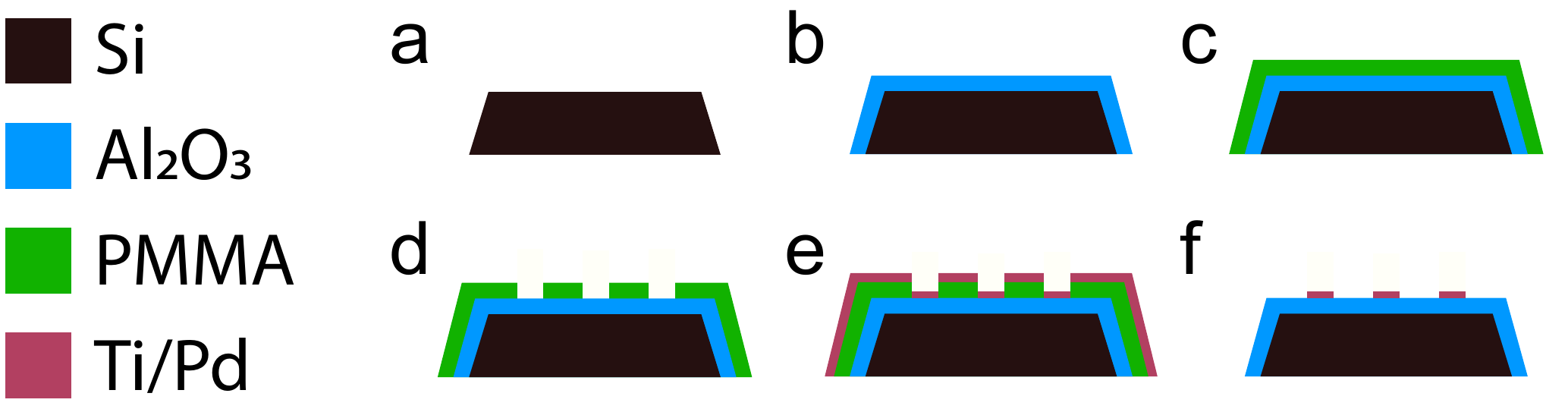}
	\caption{Protocol of gate patterning through lift-off process. (a) Side view of the prepatterned flat mesa on a cantilever tip, (b) mesa is covered with dielectric layer using ALD, (c) floating PMMA transfer, (d) definition of gate pattern through EBL, (e) electron beam evaporation of Ti/Pd, (f) lift-off yields the desired nanostructures.
	}
	\label{fig:Figure2}
\end{figure}

\begin{figure*}[t]
	\includegraphics[width=0.96\textwidth]{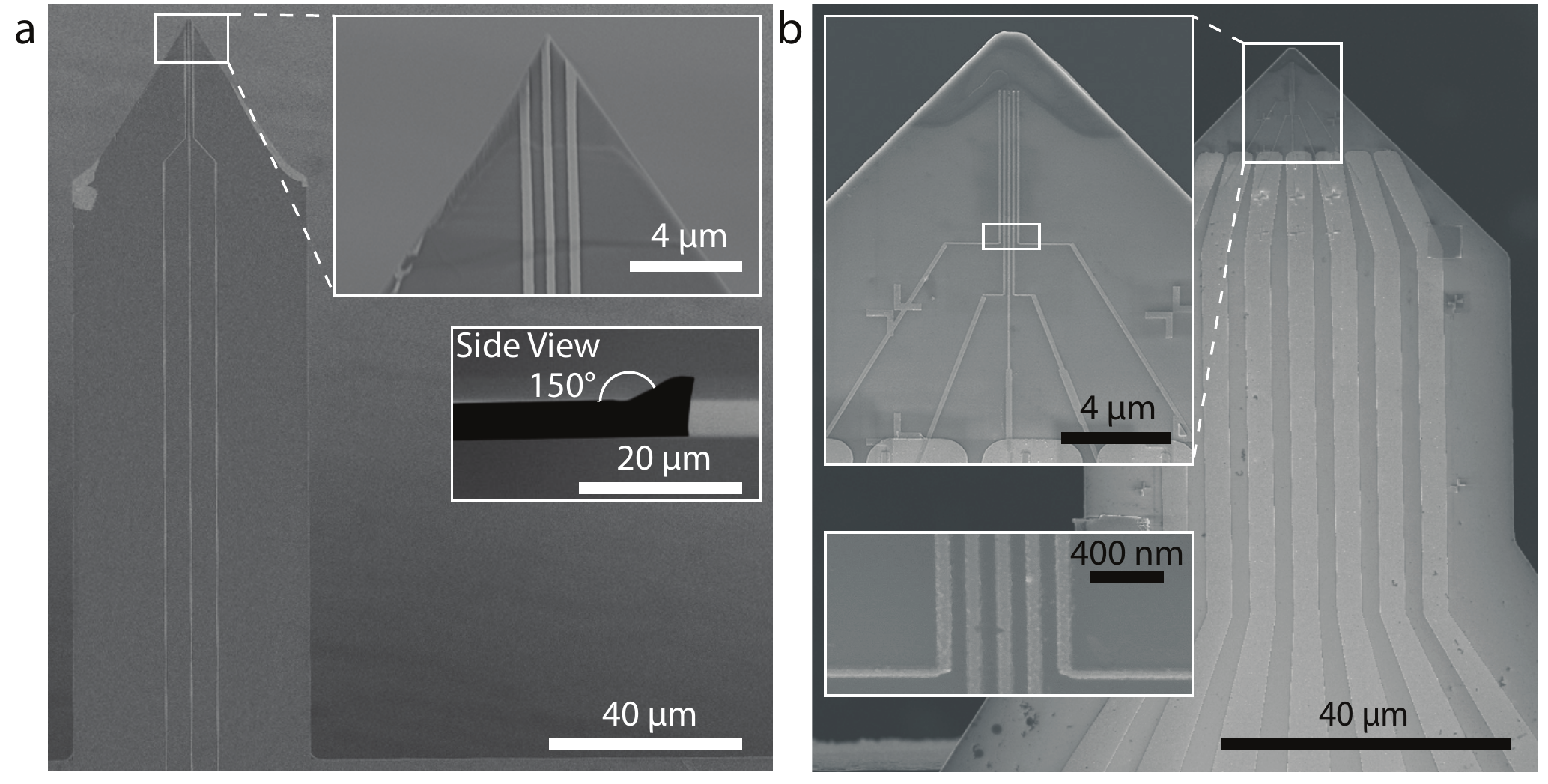}
	\caption{Scanning electron micrographs of cantilevers with gates produced through lift-off process. (a) Tipped cantilever with three Ti/Pd gates. Insets: zoom-in of tip area (top) and side view of cantilever tip part (bottom). (b) Flat cantilever with seven large Au gates connected to five finer Ti/Pd gates in the front part of the cantilever. Insets: zoom-ins of front of cantilever, showing finer gates. Note that the marker structures seen here are deposited together with the finer gates, and are used for subsequent lithography not discussed here. Bottom inset is zoom-in of area delineated by white box in top inset. Note that focussing of the electron beam on specific cantilever parts for alignment purposes results in selective additional metallization, such as the rectangular piece on the right side of the cantilever in b).
	}
	\label{fig:Figure3}
\end{figure*}

\noindent In order to define gates using the lift-off process, we first structure the non-contact AFM cantilever using focused ion beam (FIB) milling. Originally, the slope of the AFM tip with respect to the main cantilever surface is \SI{54.7}{\degree}, resulting from the anisotropic KOH etching of silicon used to fabricate the cantilevers. We succesfully cover cantilevers with such large slope angles with PMMA and perform EBL showing good pattern transfer after development. However, with these large slope angles, we find that the lift-off process does not fully remove the unexposed parts of the resist in the tip area. To improve lift-off, we therefore prestructure the tip by milling away part of it, adjusting its slope to $\sim$\SI{30}{\degree} (see inset \cref{fig:Figure3}a). We also mill away selective parts of the tip in this step, resulting in a flat mesa at the top of the tip (see \cref{fig:Figure2}a), which is used as a platform to host a set of parallel gate electrodes. Note that structuring the cantilever tip area can be done already during cantilever fabrication, allowing to mass-produce cantilevers with the required tip geometry.

\noindent Before fabricating the gates, we use atomic layer deposition (ALD) to cover the cantilever with a \SI{60}{\nano\meter} thick layer of Al$_2$O$_3$ (\cref{fig:Figure2}b), which later acts as an insulating layer between the metallic gates and the highly doped silicon cantilever body. Using the floating-layer method described before, we then coat cantilevers with PMMA (\cref{fig:Figure2}c). Here, we coat the cantilevers with three layers of 950K PMMA resist, each spun at \SI{4000}{rpm} for \SI{40}{\second} and baked at \SI{150}{\degreeCelsius} for \SI{3}{\minute} leading to a total resist thickness of roughly \SI{690}{\nano\meter}. We align on characteristic points of the cantilevers to define the markers (Ti/Au, \SI{5}{\nano\meter}/\SI{50}{\nano\meter}) in a first standard EBL step. A second EBL step is employed to define the gate patterns (\cref{fig:Figure2}d), using the predefined markers. The resist is exposed using an acceleration voltage of \SI{30}{kV} and a dose of \SI{375}{\micro\coulomb\per\centi\meter\squared}, followed by development in isopropyl alcohol (IPA) : methyl isobutyl ketone (MIBK) (3:1) for \SI{60}{\second} and IPA for \SI{60}{\second}. Afterwards, \SI{5}{\nano\meter} Ti and \SI{25}{\nano\meter} Pd are deposited using an electron beam evaporator (\cref{fig:Figure2}e). For lift-off, the sample is submerged in acetone for at least \SI{1}{h} at \SI{50}{\degreeCelsius} (\cref{fig:Figure2}f).

\noindent \cref{fig:Figure3} shows scanning electron micrographs of cantilevers patterned with gates using our method. In \cref{fig:Figure3}a, the tip is modified using FIB milling, as described. Three gates with a width of \SI{350}{\nano\meter} and pitch of \SI{400}{\nano\meter} are fabricated on the cantilever, including on the sloped tip area. For \cref{fig:Figure3}b, we use one layer of resist (\SI{230}{\nano\meter}) and pattern five nanoscale gates (width: \SI{70}{\nano\meter}, pitch: \SI{90}{\nano\meter}) on a flat cantilever. The nanoscale gates are patterned such that they connect to larger, prefabricated, contacts which widen out to bondpads on the cantilever base. 
\section*{Gates on cantilevers - etching process}
\begin{figure}[h]
	\includegraphics[width=0.48\textwidth]{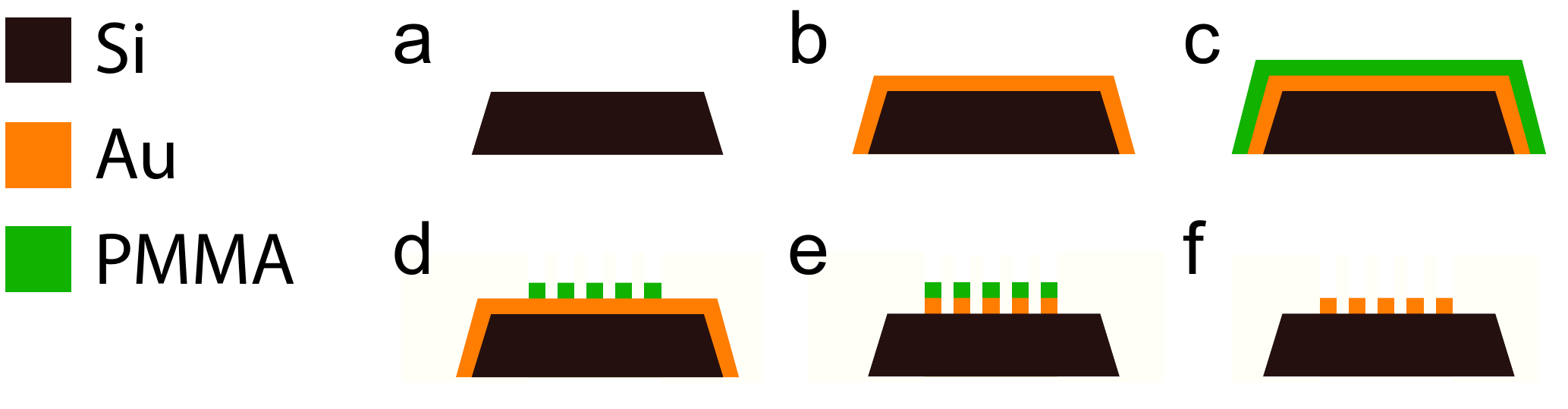}
	\caption{Protocol of gate patterning through etching process. (a) Side view of the prepatterned flat mesa on a cantilever tip, (b) electron beam evaporation of Au, (c) floating PMMA transfer, (d) gates are patterned using overdosing of PMMA, (e) argon ion beam etching transfers the pattern into the Au layer, (f) the resist is removed using oxygen reactive ion etching, yielding the desired nanostructures.
	}
	\label{fig:Figure4}
\end{figure}

\begin{figure}[h]
	\includegraphics[width=0.48\textwidth]{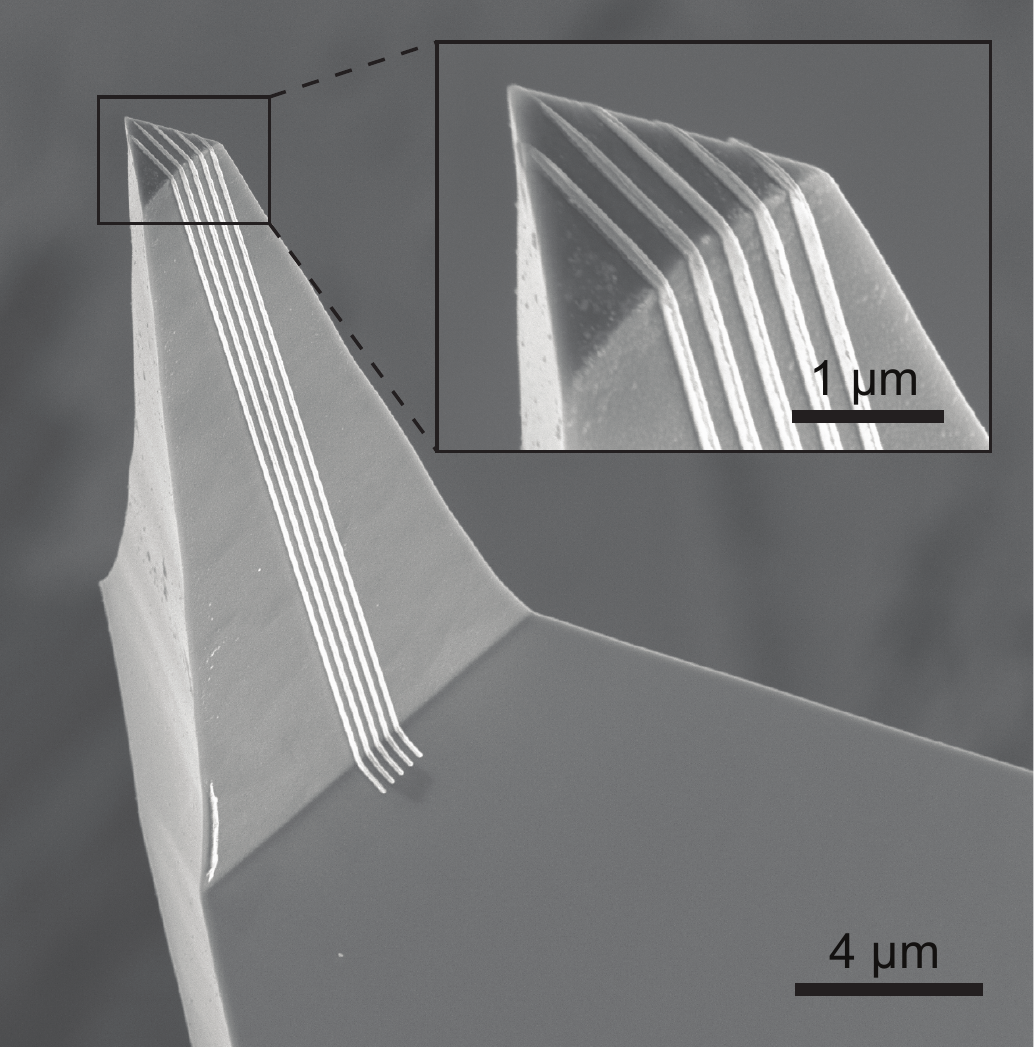}
	\caption{Scanning electron micrograph of tipped end of cantilever, patterned with gate array defined through etching process.
	}
	\label{fig:Figure5}
\end{figure}

\noindent In order to be able to fabricate gates on cantilever tips with higher slopes, we developed an etching process which does not require a lift-off step, thus avoiding problems due to incomplete material lift-off. By overdosing the same type of PMMA as before, it can be used as a negative resist~\cite{Cai2015} and act as a mask for pattern transfer through subsequent etching. This process is therefore complementary to the one described before and more suitable for removing large parts of a metallic film.

\noindent \cref{fig:Figure4} schematically shows our top-down fabrication strategy to fabricate on a steep slope, as for example on the \SI{54.7}{\degree} slope of commercially available tipped cantilevers. 
As before, we start with such a cantilever where a mesa is cut using FIB (\cref{fig:Figure4}a), to host the nanoscale gates at the end of the cantilever tip. Next, \SI{20}{\nano\meter} of Au is deposited using an electron-beam evaporator (\cref{fig:Figure4}b). As shown in the work of Cai et al.~\cite{Cai2015}, overdosing a conventionally positive PMMA resist allows to use it as a negative resist. To coat our lever, we use the technique described in \cref{fig:Figure1} (\cref{fig:Figure4}c). Next, the PMMA layer is exposed to electron beam radiation at a high dose of \SI{80000}{\micro\coulomb\per\centi\meter\squared} using an acceleration voltage of \SI{30}{\kilo\volt}. Afterwards, the sample is developed in acetone for \SI{10}{\minute} and in IPA for \SI{2}{\minute} (\cref{fig:Figure4}d). Then, argon ion beam etching (\SI{60}{\second} at \SI{0}{\degree} angle and \SI{60}{\second} at \SI{40}{\degree} angle, \SI{500}{\volt}, \SI{20}{\milli\ampere}) is used to transfer the pattern of the negative PMMA resist to the Au film (\cref{fig:Figure4}e). The last step of the process is to remove the overexposed PMMA, which is done by using oxygen reactive ion etching for \SI{10}{\minute} at \SI{120}{\watt} (\cref{fig:Figure4}f). \cref{fig:Figure5} shows a scanning electron micrograph of a cantilever patterned using the described etching process. Five gates of \SI{100}{\nano\meter} width and \SI{180}{\nano\meter} pitch are fabricated on the tipped end of the cantilever featuring the original \SI{54.7}{\degree} slope.

\section*{Mechanical properties}
\noindent For the cantilevers to maintain their function as sensitive scanning probe force transducers, it is important that their mechanical properties remain unaffected by the lithography processes. To evaluate this, we measure (in high vacuum, at room temperature) the resonance frequency and quality factor of the first flexural mode of a cantilever, before and after processing. We measure the quality factor using a ringdown method. 
Before processing, we measure a tipped cantilever with a quality factor of \SI{80}{k} and a resonance frequency of \SI{306.8}{kHz}. Next, we test the effect of our floating PMMA coating and lift-off process, first without any metal deposition. Here, we find no significant change in resonance frequency or quality factor.
Finally, we measure after performing the lift-off process, this time with deposition of \SI{5}{\nano\meter} Ti and \SI{25}{\nano\meter} Pd gates, as in \cref{fig:Figure3}a. Now, the resonance frequency is decreased to \SI{301.4}{kHz} and the quality factor to \SI{70}{k}. The decrease of the resonance frequency is expected due to the higher mass of the cantilever. The decrease in quality factor results in a modest decrease of thermal force sensitivity by $<$\SI{15}{\percent}.

\section*{Outlook}
\noindent In future work, the resolution of the gate patterns could be improved by the use of thinner PMMA layers, or by etching down the developed PMMA layer before pattern transfer~\cite{Cai2015}. An interesting use of our method would be to pattern superconducting devices, such as superconducting quantum interference devices~\cite{Wyss2022} or charge qubits~\cite{Jaeck2020} of small dimensions, on cantilever tips. This would enable sensitive imaging of nanoscale magnetic fields and charge dynamics with high resolution, for instance in 2D materials~\cite{Marchiori2021} or qubit devices~\cite{Marchiori2022}. Finally, although not tested here, our method could be used to perform EBL on various other types of samples with irregular surfaces, e.g. to pattern nanostructures on the facets of cleaved optical fibers.

\section*{Acknowledgements}
\noindent We thank Ute Drechsler, Nadine Leisgang, and Marcus Wyss for helpful discussions and technical support. We acknowledge the support of the Swiss National Science Foundation via the NCCR SPIN, the Canton Aargau, and the European Commission under H2020 FET Open grant "FIBsuperProbes" (Grant No. 892427).
 

\newpage
\begin{thebibliography}{200}
\bibitem{Meyer2021} E. Meyer, R. Bennewitz, H. Hug, \textit{Scanning probe microscopy, Second edition} (Springer, 2021).

\bibitem{Giessibl2003} F. Giessibl, \textit{Rev. Mod. Phys.} \textbf{75}, 949 (2003).

\bibitem{Wyss2022} M. Wyss, K. Bagani, D. Jetter, E. Marchiori, A. Vervelaki, B. Gross, J. Ridderbos, S. Gliga, C. Sch\"onenberger, and M. Poggio, \textit{Phys. Rev. Appl.} \textbf{17}, 034002 (2022).

\bibitem{Khivrich2020} I. Khivrich and S. Ilani, \textit{Nat. Commun.} \textbf{11}, 2299 (2020).

\bibitem{Yoo1997} M. Yoo, T. Fulton, H. Hess, R. Willett, L. Dunkleberger, R. Chichester, L. Pfeiffer, and K. West,  \textit{Science} \textbf{276}, 579 (1997).

\bibitem{Vieu2000} C. Vieu, F. Carcenac, A. Pepin, Y. Chen, M. Mejias, A. Lebib, L. Manin-Ferlazzo, L. Couraud, H. Launois, \textit{Appl. Surf. Sci.} \textbf{164}, 111 (2000).

\bibitem{Linden2011} J. Linden, C. Thanner, B. Schaaf, S. Wolff, B. L\"agel, and E. Oesterschulze, \textit{Microelectron. Eng.} \textbf{88}, 2030 (2011).

\bibitem{Han2012} A. Han, A. Kuan, J. Golovchenko, D. Branton, \textit{Nano Lett.} \textbf{12}, 1018 (2012).

\bibitem{Zhang2014} J. Zhang, C. Con, and B. Cui \textit{ACS Nano} \textbf{8}, 3483 (2014).

\bibitem{Nilsen2019} M. Nilsen, F. Port, M. Roos, K. Gottschalk, and S. Strehle, \textit{J. Micromech. Microeng.} \textbf{29}, 025014 (2019).

\bibitem{Chang2014} J. Chang, Q. Zhou, and A. Zettl, \textit{Appl. Phys. Lett.} \textbf{105}, 173109 (2014).

\bibitem{Zhou2000} H. Zhou, B. K. Chong, P. Stopford, G. Mills, A. Midha, L. Donaldson, and J. M. R. Weaver, \textit{J. Vac. Sci. Technol. B} \textbf{18}, 3594 (2000).

\bibitem{Jaafar2020} M. Jaafar, J. Pablo-Navarra, E. Berganza, P. Ares, C. Mag\'en, A. Masseboeuf, C. Gatel, E. Snoeck, J. G\'omez-Herrero, J. de Teresa, and A. Asenjo, \textit{Nanoscale} \textbf{12}, 10090 (2020).

\bibitem{Leis2021} N. M. Leisgang, \textit{Doctoral dissertation} (2021).

\bibitem{Cai2015} H. Cai, K. Zhang, and X. Yu, \textit{AIP Adv.} \textbf{5}, 117216 (2015).

\bibitem{Jaeck2020} B. J\"ack, \textit{Phys. Rev. Research} \textbf{2}, 043031 (2020).

\bibitem{Marchiori2021} E. Marchiori, L. Ceccarelli. N. Rossi, L. Lorenzelli, C. Degen, and M. Poggio, \textit{Nat. Rev. Phys.} \textbf{4}, 49 (2021).

\bibitem{Marchiori2022} E. Marchiori, L. Ceccarelli. N. Rossi, G. Romagnoli, J. Herrmann, J. Besse, S. Krinner, A. Wallraff, and M. Poggio,, \textit{Appl. Phys. Lett.} \textbf{121}, 052601 (2022).

\end{thebibliography}
\end{document}